\documentclass[epj]{svjour}
\usepackage{graphicx}
\usepackage{amsmath}
\usepackage{bm}
\usepackage{color}

\newcommand{\bald}[1]{{\bf #1}}

\usepackage{dcolumn}
\usepackage{bm}
\usepackage{graphics}
\begin{document}
\authorrunning
\titlerunning
\title{Comparing different freeze-out scenarios in azimuthal hadron correlations induced by fast partons}

\author{R. Bryon Neufeld\inst{1}
}                     
\institute{Duke University, Durham, NC 27708}
\date{Received: date / Revised version: date}
%
\abstract{
I review the linearized hydrodynamical treatment of a fast parton traversing a perturbative quark-gluon plasma.  Using numerical solutions for the medium's response to the fast parton, I obtain the medium's distribution function which is then used in a Cooper-Frye freeze-out prescription to obtain an azimuthal particle spectrum.  Two different freeze-out scenarios are considered which yield significantly different results.  I conclude that any meaningful comparison of azimuthal hadron correlation functions to RHIC data requires implementing a realistic freeze-out scenario in an expanding medium.
} 
\maketitle
\section{Introduction}
\label{intro}
An exciting result from heavy-ion experiments performed at the Relativistic Heavy-Ion Collider (RHIC) is the creation of a state of deconfined quarks and gluons known as the {\it quark-gluon plasma} (QGP) (see, for example, \cite{Adcox:2004mh}).  Fast partons (partons which propagate through the medium with a velocity approaching the speed of light) are created by hard transverse scattering in the early moments of a heavy-ion collision and are a useful probe of QGP physics.  As a fast parton propagates through the medium it interacts with the surrounding plasma, depositing energy and momentum through the process of {\it jet quenching} (see, e.~g., \cite{Baier:2000mf,Jacobs:2004qv}).  A natural question arises: how does the energy and momentum deposited by a fast parton affect the bulk behavior of an evolving QGP?

This question has become particularly interesting in light of the results of experimental measurements at RHIC \cite{Adams:2005ph,Adler:2005ee,Ulery:2007zb,:2008cq} of hadron correlation functions which suggest the fast parton may produce a propagating Mach cone in the medium.  The hadron correlation function is a measurement of the azimuthal dependence of hadrons associated with a high energy trigger hadron.  For instance, two partons may experience hard back to back transverse scattering near the surface of the medium.  One parton will rapidly escape the medium and decay into a jet of hadrons, one of which is measured and serves as the {\it near side} trigger.  The second parton propagates through the medium in the direction of the {\it away side}, depositing energy and momentum.  The hadron correlation functions show a double peak structure in the away side distribution which is consistent with the formation of a Mach cone-shaped emission pattern.

Several authors have done a theoretical investigation into how the energy and momentum deposited by the fast parton affects the bulk behavior of an evolving QGP (see, for example, \cite{CasalderreySolana:2004qm,Stoecker:2004qu,Chaudhuri:2005vc,Satarov:2005mv,Ruppert:2005uz,Renk:2005si,Friess:2006fk,Chesler:2007sv,Betz:2008js,Neufeld:2008fi,Neufeld:2008hs,Neufeld:2008dx}).
In light of the strong evidence \cite{Ludlam:2005gx,Gyulassy:2004zy} that the QGP produced at RHIC obeys the hydrodynamic assumption of local thermal equilibrium, the common theoretical approach has been to treat the fast parton as a source of energy and momentum coupled to the hydrodynamic equations of the medium.  This type of analysis requires deriving a form for the hydrodynamic source term generated by the fast parton.  In a recent work, Betz {\it et al.} \cite{Betz:2008wy} made a direct comparison of two hydrodynamic source terms, one derived in the context of perturbative quantum chromodynamics (pQCD) \cite{Neufeld:2008hs}, and the other calculated in the strongly coupled Ads/CFT correspondence \cite{Chesler:2007sv}.  In their comparison, the authors solved for the hydrodynamic evolution of an infinitely large, static background medium in the presence of the source terms.  They then obtained an azimuthal away side associated hadron correlation spectrum after performing an isochronous Cooper-Frye (CF) freeze-out.  The authors found propagating Mach cones in the hydrodynamic evolution resulting from both source terms; however, in the final hadron correlation spectrum the conical Mach structure was washed out by the CF freeze-out scenario, in disagreement with the experimental data.  An exception was found in flow generated by the non-equilibrium Neck zone in the Ads/CFT case, which allows for a conical structure in the final hadron correlation spectrum; however, this contribution does not obey Mach's law.

While the work discussed above allows for an interesting comparison of two different source terms in a particular setting, one has to question whether an isochronous CF freeze-out provides an accurate description of the hadronization that occurs in heavy-ion collisions.  An isochronous, or fixed time, freeze-out assumes that all of the matter hadronizes at the same time.  The freeze-out prescription requires integrating the distribution function over the freeze-out hypersurface, which, for an isochronous freeze-out, is just the entire volume of matter.  This prescription is particularly questionable for the case of a fast parton propagating through the QGP at RHIC, where it is likely that a Mach shock would push matter out of the medium.  This matter would then hadronize before the rest of the matter in the medium.  The effect of integrating the distribution function over the entire volume of matter at a fixed time, is that any cylindrically symmetric, or conical, contributions tend to be washed out.  In what follows, I will examine the azimuthal associated hadron correlation spectrum resulting from the hydrodynamic evolution generated in the presence of the pQCD source term derived in \cite{Neufeld:2008hs}.  It will be shown that the final spectrum is sensitive to the specific freeze-out assumptions employed.  I conclude that any meaningful comparison with experimental azimuthal hadron correlation functions measured at RHIC requires implementing a realistic freeze-out scenario in an expanding medium.

\section{Hydrodynamic Evolution}
\label{sec:1}

I now consider the hydrodynamical equations for a system in the presence of a source term, $J^\nu$, which are
\begin{equation}\label{sourcehydro}
\partial_\mu T^{\mu \nu} = J^{\nu}
\end{equation}
where $T^{\mu \nu}$ is the energy-momentum tensor of the system.  The equations can be linearized provided that the energy and momentum density deposited by the fast parton is small compared to the equilibrium energy density of the medium.  In this case one can write $T^{\mu\nu} = T_0^{\mu\nu} + \delta T^{\mu\nu}$, where $\delta T^{\mu\nu}$ is the perturbation of the energy-momentum tensor resulting from the source in an otherwise static medium.  Eq. (\ref{sourcehydro}) now takes the form
\begin{equation}\label{lin_source}
\partial_\mu \delta T^{\mu\nu} = J^\nu ,
\end{equation}
where $\partial_\mu T_0^{\mu \nu} = 0$ and $\delta T^{\mu\nu}$ is given by \cite{CasalderreySolana:2004qm}
\begin{equation}\label{mom_space_tensor}
\begin{split}
&\delta T^{00} \equiv \delta \epsilon \text{,    }\delta T^{0i} \equiv \bald{g}, \\
& \delta T^{ij} = \delta_{ij} c_s^2 \delta \epsilon - \frac{3}{4} \Gamma_s\left(\partial^i g^j + \partial^j g^i - \frac{2}{3}\delta_{ij} \bald{\nabla}\cdot\bald{g}\right).
\end{split}
\end{equation}
In (\ref{mom_space_tensor}), $\delta \epsilon$ and $\bald{g}$ are interpreted as the energy density and momentum density, respectively, deposited by the source parton, which interacts with the medium via $J^\nu$.  Also, $c_s$ denotes the speed of sound, $\Gamma_s \equiv \frac{4 \eta }{3(\epsilon_0 + p_0)} = \frac{4 \eta }{3 s T}$ is the sound attenuation length, $\epsilon_0$ and $p_0$ are the unperturbed energy density and pressure, respectively, and $\eta$ is the shear viscosity of the medium.

$\delta \epsilon$ and $\bald{g}$ must now be solved for in terms of $J^\nu$.  It is easiest to work in momentum space by taking the Fourier transform of Eq. (\ref{lin_source}) which is done using the general rule for Fourier transforms
\begin{equation}\label{generalrule}
F({\bf x},t) = \frac{1}{(2 \pi)^4}\int d^3 k \int d \omega \,e^{i \bald{k}\cdot\bald{x} - i \omega t} F({\bf k},\omega).
\end{equation}
In this case one finds \cite{Neufeld:2008dx}
\begin{eqnarray}\label{eps}
\delta\epsilon ({\mathbf k},\omega) &=& \frac{i k J_L({\mathbf k},\omega)  + J^0({\mathbf k},\omega)(i \omega -  \Gamma_s k^2)}{\omega^2 -  c_s^2 k^2 + i \Gamma_s \omega k^2} \\
\label{gl}
{\mathbf g}_L ({\mathbf k},\omega) &=& \frac{ i \omega \hat{\mathbf k} J_L({\mathbf k},\omega)+ i c_s^2 {\mathbf k} J^0({\mathbf k},\omega)}{\omega^2 -  c_s^2 k^2 + i \Gamma_s \omega k^2} \\
\label{gt}
\bald{g}_T({\mathbf k},\omega) &=& \bald{g} - \bald{g}_L = \frac{i{\mathbf J}_T({\mathbf k},\omega)}{\omega +  \frac{3}{4}i \Gamma_s k^2}
\end{eqnarray}
where the source and perturbed momentum density vectors are divided into transverse and longitudinal parts: ${\mathbf g} = \hat{\mathbf k} g_L + {\mathbf g}_T$ and ${\mathbf J} = \hat{\mathbf k} J_L + {\mathbf J}_T$, with $\hat{\mathbf k}$ being the unit vector in the direction of ${\mathbf k}$.  Eq. (\ref{gt}) is a diffusion equation and does not describe sound propagation.  The quantity ${\mathbf g}_T$ gives the diffusive momentum density generated by the fast parton.  Eqs.~(\ref{eps}) and (\ref{gl}), on the other hand, are damped wave equations.  They describe damped sound waves propagating at speed $c_s$.  Conical Mach cone structure in the medium dynamics are generated by $\delta\epsilon$ and ${\mathbf g}_L$.

The medium's dynamics can now be solved for by inserting the explicit form of $J^\nu$ in Eqs. (\ref{eps})-(\ref{gt}) and Fourier transforming back to position space.  As mentioned above, I will work with the source term derived in \cite{Neufeld:2008hs}, where the fast parton was treated as the source of an external color field interacting with a perturbative QGP in the context of kinetic theory.  In the relativistic limit ($\gamma = (1-u^2)^{-1/2} \gg 1$), the position space representation of the source term in a gluonic medium at temperature $T$ in the presence of a parton moving with velocity $\bald{u} = u \hat{x}$ at position $\bald{r} = u t\hat{x}$ is given by
\begin{equation}\label{compact}
J^\nu(x) = \left(J^0(x),{\bf u} J^0(x) - {\bf J}_\text{v}\right)
\end{equation}
where
\begin{eqnarray}\label{jnot}
J^0(\rho,z,t) &=& d(\rho,x,t) \gamma u^2 \left( 1 - \frac{\gamma u x_-}{\sqrt{x_-^2 \gamma^2 + \rho^2}} \right) \\
{\bf J}_\text{v}(\rho,x,t) &=& \left(\bald{x} - \bald{u} t \right) d(\rho,x,t) \frac{u^4}{\sqrt{x_-^2\gamma^2 + \rho^2}}
\end{eqnarray}
and
\begin{equation}
d(\rho,x,t) = \frac{\alpha_s ({Q_p^a})^2 m_{\rm D}^2}{8\pi(\rho^2 + \gamma^2 x_-^2)^{3/2}}.
\end{equation}
In the above expressions, $\rho = (z^2 + y^2)^{1/2}$ is the radius transverse to the $x$ axis, $({Q_p^a})^2 = 3$ if the source parton is a gluon and 4/3 for a quark, $\alpha_s = g^2/4\pi$ is the strong coupling, $x_- = (x - u t)$, and $m_{\rm D} = gT$ is equal to the Debye mass for gluons in the hard thermal loop approximation.

The solutions for $\delta \epsilon(\bald{x},t)$ and $\bald{g}(\bald{x},t)$ in the presence of the source term, (\ref{compact}), for a wide range of parameters have been presented in \cite{Neufeld:2008fi}, and in greater detail in \cite{Neufeld:2008dx}.  The resulting dynamics have been shown to contain a propagating Mach cone and diffusive wake.  My purpose here is not to revisit those results, but rather to use them to construct a distribution function, which can then be used in the CF freeze-out prescription to obtain an azimuthal final particle spectrum.

Given solutions for $\delta \epsilon(\bald{x},t)$ and $\bald{g}(\bald{x},t)$, one can determine the induced flow velocity and temperature modification in the medium.  The ideal hydrodynamic energy momentum tensor is given by
\begin{equation}
T^{\mu\nu} = T_0^{\mu\nu} + \delta T^{\mu\nu} = (\epsilon + p)u^\mu u^\nu - p g^{\mu\nu}
\end{equation}
where $u^{\mu}$ is the flow velocity of the medium, which for the unperturbed static medium is just given by $u^\mu = (1,0,0,0)$.  Thus, working to first order in perturbed quantities, one has for $T^{0i}$
\begin{equation}
\begin{split}\label{delu}
T^{0i} &= T_0^{0i} + \bald{g} = \\
&(\epsilon_0 + \delta \epsilon + p_0 + \delta p)(u_0 + \delta u)^0 (u_0 + \delta u)^i \\
&= T_0^{0i} + (\epsilon_0 + p_0) \delta \bald{u}.
\end{split}
\end{equation}
I conclude that
\begin{equation}
\delta\bald{u} = \frac{\bald{g}}{\epsilon_0 + p_0} = \frac{3 \bald{g}}{4 \epsilon_0}
\end{equation}
where I have used $\epsilon_0 = 3 p_0$. One can express $\delta u^0$ in terms of $\delta \bald{u}$ by requiring the medium's flow velocity remains physical, or $u^\mu u_\mu = 1$, which eventually yields
\begin{equation}
\delta u_0 = \frac{\bald{u}\cdot\delta \bald{u}}{{u_0}_0}.
\end{equation}
An expression for $\delta T$ can similarly be found by using the equation of state for an ideal gas of gluons
\begin{equation}
\epsilon = \frac{8 \pi^2\, T^4}{15},
\end{equation}
in which case one obtains
\begin{equation}\label{delT}
\delta T = \frac{\delta \epsilon}{4 \epsilon_0} T_0.
\end{equation}

\section{Freeze-out}
\label{sec:2}

Having expressions for the flow velocity and temperature, it is now possible to construct the medium's distribution function, which in the Boltzmann limit, and up to normalization, is given by
\begin{equation}
f(\bald{r},\bald{p},t) = e^{-\beta u^\mu p_\mu} = \exp{\left[-\frac{(u_0 + \delta u)^\mu p_\mu}{T_0 + \delta T}\right]}.
\end{equation}
As discussed above, the distribution is converted into an azimuthal particle distribution by using a Cooper-Frye freeze-out scenario \cite{Cooper:1974mv}.  Consistent with the approach discussed in \cite{Betz:2008wy,Noronha:2008tg}, the final azimuthal particle spectrum for massless particles at mid-rapidity $(y = 0)$ is given by
\begin{equation}\label{cfform}
\frac{d N}{dy \, d\phi}(y = 0) = \int_{p^i_T}^{p^f_T} d p_T \, p_T \int d \Sigma_\mu P^\mu (f(\bald{r},\bald{p},t) - f_{0})
\end{equation}
where $\Sigma_\mu$ is the freeze-out hypersurface and
\begin{equation}
P^\mu = (p_T, p_T \cos(\pi - \phi), p_T\sin(\pi - \phi),0).
\end{equation}
The isotropic background contribution, $f_0 = e^{-\beta_0 u_0^\mu p_\mu}$, is subtracted in (\ref{cfform}).  In what follows, it is understood that the source parton propagates along the $-\hat{x}$ axis.

I now consider the spectrum given by (\ref{cfform}), using the results for $\delta \epsilon(\bald{x},t)$ and $\bald{g}(\bald{x},t)$ which are discussed in \cite{Neufeld:2008fi}.  The following parameters are used: $\alpha_s = 1/\pi$, $c_s = 0.57$, $\eta/s = 1/4\pi$, $\gamma = 33$, $({Q_p^a})^2 = 3$ and $T = 350$ MeV.  For the sake of clarity, in what follows I will not include the diffusive contribution to $\bald{g}$ from $\bald{g}_T$.  This will ensure that all contributions to the final particle spectrum come from the conical Mach cone structure generated by $\delta\epsilon$ and ${\mathbf g}_L$.  I have also added a Gaussian ansatz to the distribution centered around $\phi = 0$ to mimic the near side, or trigger, contribution to the hadron correlation functions observed at RHIC.  The physical contribution, that obtained from the solutions for $\delta \epsilon(\bald{x},t)$ and $\bald{g}_L(\bald{x},t)$, will be centered around $\phi = \pi$.

\begin{figure}
\centerline{
\includegraphics[width = 0.85\linewidth]{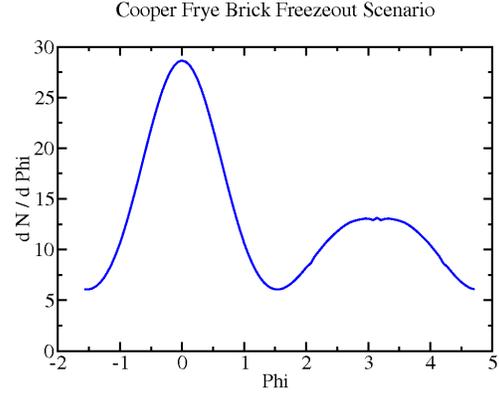}
}
\caption{(Color online) The azimuthal particle spectrum generated by a fast parton in an isochronous freeze-out scenario (as discussed in the text).  The peak centered at $\phi = 0$ has been added to simulate the near side distribution observed at RHIC, while the away side peak centered at $\phi = \pi$ results from the medium dynamics.   Although the medium response considered here is purely conical in structure, the away side peak is a broad Gaussian rather than double peaked.  The isochronous CF freeze-out scenario has washed out any evidence of the conical structure present in the medium's dynamics.
}
\label{brick}
\end{figure}

\begin{figure}
\centerline{
\includegraphics[width = 0.85\linewidth]{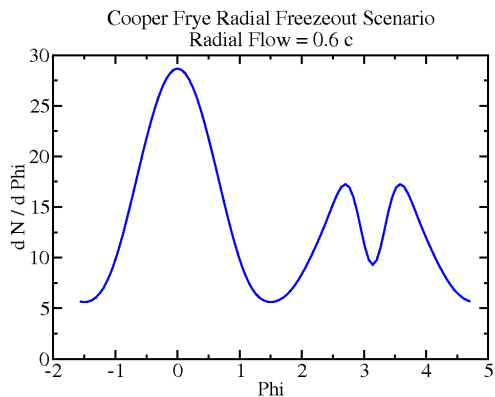}
}
\caption{(Color online) The azimuthal particle spectrum generated by a fast parton in a radial shell freeze-out scenario with artificially added background velocity (as discussed in the text).  The away side spectrum contains a markedly different structure than that seen in Fig. \ref{brick}.  In particular, the away side distribution now contains a double peak structure, reflecting the conical flow present in the medium.
}
\label{radflow}
\end{figure}

I first consider an isochronous freeze-out scenario, similar to that done in \cite{Betz:2008wy}.  In this scenario, the hypersurface is given by $d \Sigma_\mu = (d V,0,0,0)$.  The matter is assumed to instantly convert into free streaming particles, and the hypersurface integration in (\ref{cfform}) becomes an integration over all space.  For the momentum integration in (\ref{cfform}) I have chosen $p^i_T = 4$ GeV and $p^f_T = 8$ GeV.  The result is shown in Fig. \ref{brick} for an arbitrary normalization.  As mentioned above, the contribution around $\phi = 0$ simulates the near side distribution and has been added to provide context.  The away side distribution is centered around $\phi = \pi$ and is generated by the solution for $\delta \epsilon(\bald{x},t)$ and $\bald{g}_L(\bald{x},t)$.  Even though $\delta \epsilon(\bald{x},t)$ and $\bald{g}_L(\bald{x},t)$ only contribute a conical Mach cone structure to the medium's dynamics, there is no sign of this in the away side distribution, which has a broad Gaussian shape instead of a double peak structure.  The isochronous CF freeze-out scenario has washed out any evidence of the conical structure present in the medium's dynamics.

I next consider a scenario in which the freeze-out surface is a radial shell extending from $\rho = 4-5$ fm.  The source parton is assumed to start just inside the shell and then propagate through it until any contribution from the medium's hydrodynamic response is negligible.  Recall that $\rho = (z^2 + y^2)^{1/2}$ and the source parton is propagating along the $-\hat{x}$ axis.  The hypersurface is now oriented in the positive $\hat{\rho}$ direction and the integration includes time.  I limit the integration to a thin slice in the $\hat{z}$ direction, extending from $\pm 0.1$ fm.  Finally, an artificial radial background flow velocity of $v_\rho = 0.6$ is introduced in the medium velocity appearing in (\ref{cfform}).  The result is presented in Fig. \ref{radflow}, where again, the normalization is arbitrary.  As one can see, under this freeze-out scenario the away side distribution now contains a double peak structure, reflecting the conical flow present in the medium.

\section{Discussion and Conclusions}
\label{sec:3}

The results presented above show that theoretical calculations of the azimuthal away side hadron correlation spectrum generated by fast partons are sensitive to the specific freeze-out scenario employed.  Specifically, it was shown that an isochronous Cooper-Frye freeze-out scenario yielded significantly different results than a radial shell scenario with an artificially added background velocity (as discussed in Sec. \ref{sec:2}).  The results, as plotted in Figs. \ref{brick} and \ref{radflow}, are not meant to provide a meaningful comparison with the data observed at RHIC, but are rather shown to highlight the sensitivity of results to the freeze-out scenario chosen.

Indeed, even the result presented in Fig. \ref{radflow} is sensitive to the choice of parameters.  A change in the speed of sound in the medium from $c_s = 0.57$ to $c_s = 0.3$ has the effect of wiping out the double peak structure, leaving a single Gaussian peak on the away side.  A similar effect is observed if one changes $\eta/s$ from $1/4\pi$ to $6/4\pi$ while leaving the speed of sound unchanged.

The potential for ambiguity in various freeze-out scenarios is not surprising.  Freeze-out occurs in physical systems when the mean free path goes to infinity and particles become free-streaming.  This occurs in heavy-ion collisions because the medium is expanding and cooling.  A conventional choice for the freeze-out temperature is 110 MeV \cite{Nonaka:2006yn}, well below the initial temperature of the medium (around 400 MeV) and the critical temperature for hadronization (around 160 MeV).  An infinite static background medium, such as that considered above and in the comparison done in \cite{Betz:2008wy}, does not expand or cool, hence there is no reason for freeze-out to occur.  Freeze-out must be imposed on the system independent of physical processes, creating the potential for widely varying and inaccurate results.  Any meaningful theoretical comparison of azimuthal hadron correlation functions measured at RHIC requires implementing a realistic freeze-out scenario in an expanding medium.

\end{document}